\documentclass[12pt]{iopart}

\usepackage{iopams}
\usepackage{cite}

\usepackage{color}
\usepackage{graphicx}
\usepackage{mathtext}

\begin{document}

\title[Circular Cumulants Approach for Multiplicative Intrinsic Noise]{Circular Cumulant Reductions for Macroscopic Dynamics of Kuramoto Ensemble with Multiplicative Intrinsic Noise}

\author{Denis S Goldobin$^{1,2}$ and Anastasiya V Dolmatova$^{1,3}$}
\address{$^1$Institute of Continuous Media Mechanics, Ural Branch of RAS, Acad.\ Korolev street 1, 614013 Perm, Russia}
\address{$^2$Department of Theoretical Physics, Perm State University, Bukirev street 15, 614990 Perm, Russia}
\address{$^3$Institute for Information Transmission Problems, RAS, B.\ Karetny per.\ 19, 127051 Moscow, Russia}
\ead{Denis.Goldobin@gmail.com}

\begin{abstract}
We demonstrate the application of the circular cumulant approach for thermodynamically large populations of phase elements, where the Ott--Antonsen properties are violated by a multiplicative intrinsic noise.
The infinite cumulant equation chain is derived for the case of a sinusoidal sensitivity of the phase to noise.
For inhomogeneous populations, a Lorentzian distribution of natural frequencies is adopted.
Two-cumulant model reductions, which serve as a generalization of the Ott--Antonsen ansatz, are reported. The accuracy of these model reductions and the macroscopic collective dynamics of the system are explored for the case of a Kuramoto-type global coupling.
The Ott--Antonsen ansatz and the Gaussian approximation are found to be not uniformly accurate for non-high frequencies.

\vspace{0.5cm}

\end{abstract}

\submitto{\it J.\ Phys.\ A: Math.\ Theor.}

\maketitle

\section{Introduction}
Significant advance in studies of patterns and waves in active media with local, non-local, and global (\textit{or} mean-field) interactions~\cite{Kanamaru-Aihara-2008,Pazo-Montbrio-2014,Montbrio-etal-2015,Omelchenko-etal-2014,Laing-2015,Laing-2016,Laing-2017,Pimenova-etal-2016,Dolmatova-etal-2017,Smirnov-etal-2017,Bolotov-etal-2018,Smirnov-etal-2018,Omelchenko-2018,Omelchenko-2019} became possible due to the application of the Ott--Antonsen theory\cite{OA-2008,OA-2009}, based on the Watanabe--Strogatz partial integrability~\cite{WS-1993,WS-1994,Pikovsky-Rosenblum-2008,Marvel-etal-2009} revealed for an important broad class of paradigmatic models in the theory of collective phenomena (e.g., Kuramoto-type models~\cite{Kuramoto-1975,Kuramoto-1984,Acebron-etal-2005,Pikovsky-Rosenblum-2015}). Generalization of the Ott--Antonsen approach to the cases of realistic systems (e.g.,~\cite{Martens-etal-2013,Totz-etal-2018,Pietras-Daffertshofer-2019}) was problematic until recently, when the circular cumulant formalism was introduced~\cite{Tyulkina-etal-2018,Goldobin-etal-2018,Tyulkina-etal-2019} and demonstrated for systems with an additive intrinsic noise.

While in~\cite{Tyulkina-etal-2018,Goldobin-etal-2018} the intrinsic noise in phase is additive, the case of multiplicative noise is of interest as well. For instance, in populations of quadratic integrate-and-fire neurons~\cite{Pazo-Montbrio-2014,Montbrio-etal-2015,Ulner-etal-2018,Volo-Torcini-2018}, an additive intrinsic noise in the membrane voltage results in a multiplicative noise for the oscillation phase variable.
In this letter, we show how the case of a multiplicative noise can be rigorously handled on the basis of the circular cumulant approach. The general cumulant equation chain is derived and macroscopic model reductions with two leading cumulants are suggested and tested for the case of Kuramoto-type coupling. Two-cumulant reductions provide the $\sigma^2$-correction to the Ott--Antonsen ansatz ($\sigma^2$ is the noise intensity). However, technically, they require smallness of higher-order cumulants which can be observed also for moderate and strong noise in some systems (e.g.,~\cite{Goldobin-etal-2018}). Constructed two-cumulant reductions serve as a generalization of the virgin Ott--Antonsen theory.

\section{Populations of Ott--Antonsen phase elements subject to multiplicative intrinsic noise}
\subsection{Fokker--Planck equation and dynamics of Kuramoto--Daido order parameters}
We consider a population of $N$ phase elements with first-harmonic coupling (\textit{or} ``sin-coupling'') subject to a multiplicative intrinsic noise:
\begin{equation}
\dot\varphi_n =\omega(t)+\mathrm{Im}(2h(t)\e^{-\mathrm{i}\varphi_n}) +\sqrt{2}\sigma\sin\varphi_n\,\xi_n(t)\,,\quad n=1,2,....,N,
\label{eq01}
\end{equation}
where $\varphi_n $ is the phase of the $n$-th element, which can be an oscillator, an excitable system, or a directional element, $\omega(t)$ and $h(t)$ are arbitrary real- and complex-valued functions of time (which should be the same for each element), $\sigma$ is the intrinsic noise strength, $\xi_n(t)$ are normalized independent $\delta$-correlated Gaussian noise signals:
$\langle\xi_n\rangle_\xi=0$ and $\langle\xi_n(t)\,\xi_m(t_1)\rangle_\xi=2\delta_{nm}\delta(t-t_1)$, $\langle\dots\rangle_\xi$ indicates averaging over noise realizations.
Stochastic equations~(\ref{eq01}) are considered in the Stratonovich sense.
For $\sigma=0$, in the thermodynamic limit $N\to\infty$, this system obeys the Ott--Antonsen theory~\cite{OA-2008,OA-2009}. The individual intrinsic noise violates the applicability conditions of the OA theory.

The $\sin\varphi_n$-modulation of the noise strength we adopt in Eq.~(\ref{eq01}) corresponds the case of a noise forcing of weakly anharmonic oscillators (e.g., see~\cite{Goldobin-2011,Goldobin-2014}) and, for instance, can be relevant for metronomes~\cite{Martens-etal-2013}. In~(\ref{eq01}) the multiplier $\sqrt{2}$ is introduced for normalization, as, in the high-frequency limit, the impact of an intrinsic noise on the macroscopic collective dynamics is controlled by the average $(2\pi)^{-1}\int_{0}^{2\pi}(\sqrt{2}\sigma\sin\varphi)^2\mathrm{d}\varphi=\sigma^2$~\cite{Goldobin-Dolmatova-2019}.

In the thermodynamic limit $N\to\infty$, the evolution of the probability density $w(\varphi,t)$ of phases is governed by the following Fokker--Planck equation:
\begin{eqnarray}
\frac{\partial w(\varphi,t)}{\partial t}
 +\frac{\partial}{\partial\varphi}\Big[\left(\omega(t) -\mathrm{i}h(t)\,\e^{-\mathrm{i}\varphi} +\mathrm{i}h^\ast(t)\,\e^{\mathrm{i}\varphi}\right)w(\varphi,t)\Big]
\nonumber\\
\qquad\qquad
=2\sigma^2\frac{\partial}{\partial\varphi}\left(\sin\varphi \frac{\partial}{\partial\varphi}\Big(\sin\varphi\,w(\varphi,t)\Big)
\right).
\label{eq02}
\end{eqnarray}
In Fourier space, where $w(\varphi,t)\equiv(2\pi)^{-1}\sum_{m=-\infty}^{+\infty}a_m(t)\e^{-\mathrm{i}m\varphi}$ with $a_0=1$ by definition and $a_{-m}=a_m^\ast$, equation~(\ref{eq02}) reads
\begin{eqnarray}
\textstyle
\dot{a}_m=m\left(\mathrm{i}\omega(t)\,a_m +h(t)\,a_{m-1} -h^\ast(t)\,a_{m+1}\right)
\nonumber\\
\qquad
\textstyle
-\sigma^2\left(m^2a_m -\frac{m(m-1)}{2}a_{m-2} -\frac{m(m+1)}{2}a_{m+2}\right),
\;\mbox{ for } m=1,2,3,...\,.
\label{eq03}
\end{eqnarray}
Notice, here population-average $a_m=\int_0^{2\pi}w(\varphi,t)\e^{\mathrm{i}m\varphi}\mathrm{d}\varphi$ are the Kuramoto--Daido order parameters~\cite{Daido-1996}.

Below we introduce the frequency nonidentity of population elements. For simplicity, we introduce it for the case of time-independent natural frequencies $\omega_n $, although for many specific physical systems the procedure can be performed also for time-dependent $\omega_n(t)$ (e.g., see\cite{Pazo-Montbrio-2014,Montbrio-etal-2015,Petkoski-Stefanovska-2012,Ulner-etal-2018,Volo-Torcini-2018}). Let us consider a Lorentzian distribution of natural frequencies $\omega_n$ peaked at $\Omega$ and with a half width at half maximum $\gamma$: $g(\omega)=\gamma/(\pi[\gamma^2+(\omega-\Omega)^2])$. In this case, one can consider subpopulations with identical values of frequency $\omega$ and write equation~(\ref{eq03}) for $a_m(\omega,t)$ of each subpopulation. Ensemble-mean order parameters are
\begin{equation}
Z_m(t)=\int_{-\infty}^{+\infty}a_m(\omega,t)\,g(\omega)\,\mathrm{d}\omega\,.
\label{eq04}
\end{equation}
Equation system~(\ref{eq03}) for $a_m$ with positive indices does not involve $a_m$ with negative indices, which are complex conjugate $a_{-m}=a_m^\ast$. Importantly, the term $\frac12\sigma^2m(m-1)a_{m-2}$, which potentially could involve $a_{-1}=a_1^\ast$ for $m=1$, vanishes at $m=1$: $\frac12\sigma^2m(m-1)a_{m-2}=0\cdot a_{-1}$. Since system~(\ref{eq03}) for the set $\{a_m(\omega),\,m=1,2,3,...\}$ is analytic for $a_m(\omega)$ as functions of $\omega$ extended to the complex plane, one can consider analytic $a_m(\omega)$ and calculate integrals~(\ref{eq04}) via residues and find $Z_m(t)=a_m(\omega+\mathrm{i}\gamma,t)$ (see, e.g.,~\cite{Crawford-1994,OA-2008} for details). Thus, we obtain from equation system~(\ref{eq03}):
\begin{eqnarray}
\textstyle
\dot{Z}_m=m\left[(\mathrm{i}\Omega-\gamma)Z_m +h\,Z_{m-1} -h^\ast Z_{m+1}\right]
\nonumber\\
\qquad
\textstyle
-\sigma^2\left(m^2Z_m -\frac{m(m-1)}{2}Z_{m-2} -\frac{m(m+1)}{2}Z_{m+2}\right)\,,
\quad m=1,2,3,...\,.
\label{eq05}
\end{eqnarray}

\subsection{Circular cumulant description of collective dynamics}
Equation system~(\ref{eq05}) for $\sigma\ne 0$ does not admit the OA ansatz $Z_m=(Z_1)^m$ (as well as the Gaussian approximation $Z_m\approx|Z_1|^{m^2-m}(Z_1)^m$~\cite{Zaks-etal-2003,Sonnenschein-etal-2013,Sonnenschein-etal-2015,Hannay-etal-2018}). The natural framework for considering the system dynamics in the vicinity of the OA manifold is the circular cumulant approach~\cite{Tyulkina-etal-2018}, which was also found to be well consistent with experimental data~\cite{Goldobin-Dolmatova-2019b}. Considering order parameters $Z_m$ as moments of $\e^{\mathrm{i}\varphi}$, one can introduce characteristic function~\cite{Gardiner-1983,Lukacs-1970}
\begin{equation}
F(k,t)\equiv\sum_{m=0}^\infty Z_m(t)\frac{k^m}{m!} \,.
\label{eq06}
\end{equation}
With characteristic function~(\ref{eq06}), one can introduce the generating function of `circular cumulants' $\kappa_m$:
\begin{equation}
\Psi(k,t)\equiv k\frac{\partial}{\partial k}\ln{F(k,t)} \equiv\sum_{m=1}^\infty\kappa_m(t)\,k^m .
\label{eq07}
\end{equation}
The definition~(\ref{eq07}) is adopted in such an `exotic' form as it provides the simplest form of cumulant equations~\cite{Tyulkina-etal-2018} and turned out to be natural for the interrelations between the Watanabe--Strogatz variables~\cite{WS-1993,WS-1994,Pikovsky-Rosenblum-2008,Marvel-etal-2009} and the circular cumulants~\cite{Goldobin-2019}. Specifically, the three first cumulants are $\kappa_1=Z_1$, $\kappa_2 =Z_2-Z_1^2$, and $\kappa_3=(Z_3-3Z_2Z_1 +2Z_1^3)/2$.

In terms of circular cumulants the OA solution $Z_m=(Z_1)^m$ acquires a
simple form: $\kappa_1=Z_1$, $\kappa_{m>1}=0$. This one-element form,
instead of an infinite series of $Z_m$, is convenient for constructing a
perturbation theory against the background of the OA solution.

For the generating function $F(k,t)=\sum\limits_{m=0}^\infty{Z_m(t)\frac{k^m}{m!}}$, equation system~(\ref{eq05}) yields $\frac{\partial}{\partial t}F(k,t)$.
\begin{itemize}
\item For the term $mZ_m$, one can notice
 $k\frac{\partial}{\partial k}F=\sum\limits_{m=0}^\infty{mZ_m\frac{k^m}{m!}}$\,;
\item for $mZ_{m-1}$, $kF=\sum\limits_{m=0}^\infty{mZ_{m-1}\frac{k^m}{m!}}$\,;
\item for $mZ_{m+1}$,
 $k\frac{\partial^2}{\partial k^2}F=\sum\limits_{m=0}^\infty{km(m-1)Z_m\frac{k^{m-2}}{m!}} =\sum\limits_{m=0}^\infty{mZ_{m+1}\frac{k^m}{m!}}$\,;
\item for $m^2Z_m$, $k\frac{\partial}{\partial k}\left({k\frac{\partial}{\partial k}F}\right) =\sum\limits_{m=0}^\infty{m^2Z_m\frac{k^m}{m!}}$\,;
\item for $m(m-1)Z_{m-2}$, $k^2F=\sum\limits_{m=0}^\infty{m(m-1)Z_{m-2}\frac{k^m}{m!}}$\,;
\item for $m(m+1)Z_{m+2}$,
$\textstyle
k\frac{\partial^2}{\partial k^2}\left(k\frac{\partial^2}{\partial k^2}F\right) =\sum\limits_{m=0}^{\infty}m(m-1)^2(m-2)Z_m\frac{k^{m-2}}{m!}=$
\\
$\sum\limits_{m=0}^\infty{m(m+1)Z_{m+2}\frac{k^m}{m!}}$\,.
\end{itemize}
Hence,
\begin{eqnarray}
\textstyle
\partial_tF=(\mathrm{i}\Omega-\gamma)k\partial_kF +hkF -h^\ast k\partial_k^2F
\nonumber\\
\textstyle
\qquad\qquad
 -\sigma^2\left(k\partial_k(k\partial_kF) -\frac{1}{2}[k^2F+k\partial_k^2(k\partial_k^2F)]\right)\,,
\label{eq08}
\end{eqnarray}
where $\partial_t$ and $\partial_k$ stand for the $t$- and $k$- partial derivatives, respectively. Employing definition~(\ref{eq07}), one can find $\partial_t\Psi=k\partial_k\frac{\partial_tF}{F}$, and equation~(\ref{eq08}) yields
\begin{eqnarray}
\textstyle
\partial_t\Psi =k\partial _k \left[(\mathrm{i}\Omega-\gamma)\Psi +hk -h^\ast\left(k\partial_k\frac{\Psi}{k}+\frac{\Psi^2}{k}\right) -\sigma^2\left(k\partial_k\Psi+\Psi^2-\frac{k^2}{2}\right.\right.
\nonumber\\
\textstyle
\qquad\qquad
 \left.\left. -\frac{(k\partial_k-1)^2(k\partial_k-2)\Psi +(2(k\partial_k)^2-6k\partial_k+5)\Psi^2-(k\partial_k\Psi)^2 +(2k\partial_k-4)\Psi^3+\Psi^4}{2k^2}\right)\right].
\label{eq09}
\end{eqnarray}
The expression in the latter term of equation~(\ref{eq09}) was checked with the ``Maple'' package for analytical calculations. With expansion~(\ref{eq07}), equation~(\ref{eq09}) yields
\begin{eqnarray}
\textstyle
\dot\kappa_m =m\bigg[(\mathrm{i}\Omega-\gamma)\kappa_m +h\delta_{1m} -h^\ast\bigg(m\kappa_{m+1} +\sum\limits_{m_1+m_2=m+1}\!\kappa_{m_1}\kappa_{m_2}\bigg)
\nonumber\\
\textstyle
\qquad
 -\sigma^2\bigg(m\kappa_m +\sum\limits_{m_1+m_2=m}\!\kappa_{m_1}\kappa_{m_2} -\frac{1}{2}\delta_{2m}-\frac{m(m+1)^2}{2}\kappa_{m+2}
\nonumber\\
\textstyle
\qquad\quad
 -\!\sum\limits_{m_1+m_2=m+2}\!\!\frac{2m^2 +2m+1-m_1m_2}{2}\kappa_{m_1}\kappa_{m_2}
 -\!\!\sum\limits_{m_1+m_2+m_3=m+2}\!\!m\kappa_{m_1}\kappa_{m_2}\kappa_{m_3}
\nonumber\\
\textstyle
\qquad\quad
 -\sum\limits_{m_1+m_2+m_3+m_4=m+2} \frac{1}{2}\kappa_{m_1}\kappa_{m_2}\kappa_{m_3}\kappa_{m_4}\bigg)\bigg]\,.
\label{eq10}
\end{eqnarray}
For the dynamics of first three cumulants:
\begin{eqnarray}
\dot\kappa_1 =(\mathrm{i}\Omega-\gamma)\kappa_1 +h-h^\ast(\kappa_2 +\kappa_1^2) -\sigma^2\left[\kappa_1-\kappa_1^3-2\kappa_3-3\kappa_1\kappa_2\right]\,,
\label{eq11}
\\
\dot\kappa_2 =2(\mathrm{i}\Omega-\gamma)\kappa_2 -4h^\ast(\kappa_3+\kappa_1\kappa_2)
\nonumber\\
\qquad\qquad\quad
-\sigma^2\left[4(1-3\kappa_1^2)\kappa_2 -(1-\kappa_1^2)^2
 -18\kappa_4  -20\kappa_1\kappa_3 -9\kappa_2^2\right]\,,
\label{eq12}
\\
\dot\kappa_3 =3(\mathrm{i}\Omega-\gamma)\kappa_3 -3h^\ast(3\kappa_4+2\kappa_1\kappa_3+\kappa_2^2)
\nonumber\\
\qquad\qquad\quad
{}-3\sigma^2\left[3\kappa_3+2\kappa_1\kappa_2-24\kappa_5 -21\kappa_1\kappa_4-19\kappa_2\kappa_3\right.
\nonumber\\
\qquad\qquad\qquad\qquad\qquad\qquad\qquad\qquad
\left.{}-9\kappa_1^2\kappa_3-9\kappa_1\kappa_2^2-2\kappa_1^3\kappa_2\right].
\label{eq13}
\end{eqnarray}
Let us asses the orders of smallness of high-order cumulants for $\kappa_1\sim1$ and weak noise $\sigma^2\ll1$, with assumption that the reference value of $\kappa_{m+1}$ is nonlarger than that of $\kappa_m$. Equation~(\ref{eq12}) dictates for small $\kappa_2 $ the order of magnitude $\kappa_2\sim\sigma^2$. Further, $\kappa_3\sim\max\{|\kappa_2|^2,\sigma^2|\kappa_2|\}$, i.e., $\kappa_3\sim\sigma^4$. Generally, one can substitute the hierarchy $\kappa_m\sim\sigma^{2(m-1)}$ and find that it is admissible.

Alternatively, if, due to some reason, $\kappa_1\sim\varepsilon\ll1$, then, according to equation~(\ref{eq12}), $\kappa_2\sim\sigma^2$; according to equation~(\ref{eq13}), $\kappa_3\sim\sigma^4$; generally, $\kappa_{m>1}\sim\sigma^{2(m-1)}$. Thus, we have the same hierarchy, as for $\kappa_1\sim 1$. The smallness of $\sigma^2$ leads to the formation of universal hierarchy of circular cumulants which allows one to use truncated cumulant equation chains for approximate description of the macroscopic population dynamics.

\section{Two-cumulant reductions for collective dynamics}
In this letter we exemplify the application of finite cumulant-based reduced models with the Kuramoto ensemble~\cite{Kuramoto-1975,Kuramoto-1984}, i.e., the case of
\begin{equation}
h=\frac{K}{2}Z_1\,,
\label{eq14}
\end{equation}
where $K$ is the coupling coefficient.

In~\cite{Goldobin-etal-2018}, the OA ansatz $Z_m=(Z_1)^m$, the Gaussian approximation $Z_m\approx|Z_1|^{m^2-m}(Z_1)^m$, and two-cumulant reductions with three possible closures for $\kappa_3 $ where considered and compared to the `exact' solutions for the case of Kuramoto ensemble and coupled active rotators. In the case of this letter for nonlarge frequency $\Omega$, we observed that the OA ansatz and Gaussian approximation often fail dramatically even for as small noise intensity as $\sigma^2=0.1$~[\footnote{In~(\ref{eq12}), the intrinsic noise creates nonzero $\kappa_2$ even for ($\kappa_1=0$)-states; two-bunch states expectedly cannot be well represented with single-peaked wrapped Cauchy and Gaussian distributions.}]. The \textit{ad hoc} closures for $\kappa_3$ and $\kappa_4$ are also not uniformly beneficial compared to the approximation $\kappa_3=\kappa_4=0$. Hence, we restrict our presentation to two reduction models:

\begin{itemize}
\item \textbf{(C20)} is a two-cumulant truncation of equation system~(\ref{eq10}), where we set $\kappa_3=\kappa_4=0$ and discard the terms which introduce contributions below the strong accuracy level of equations dictated by hierarchy $\kappa_m\sim\sigma^{2(m-1)}$:
\begin{eqnarray}
\dot{Z}=(\mathrm{i}\Omega-\gamma)Z+h-h^\ast(\kappa+Z^2)-\sigma^2Z(1-Z^2)\,,
\label{eq15}
\\
\dot\kappa=2(\mathrm{i}\Omega-\gamma)\kappa-4h^\ast Z\kappa-\sigma^2[4\kappa-(1-Z^2)^2]\,,
\label{eq16}
\end{eqnarray}
where, for the brevity of notation and the ease of comparison to the literature, we introduce notation $\kappa_1=Z$ and $\kappa_2=\kappa$. The term $4\sigma^2\kappa$ is kept in spite of its smallness, as it introduces the dissipativity into the system dynamics, which is essential in the case of $\gamma=0$ and conservative $h$.
\end{itemize}

\begin{itemize}
\item \textbf{(C2e)} is an enhanced two-cumulant truncation of equation system~(\ref{eq10}), where we set $\kappa_3=\kappa_4=0$ but do not discard any other terms in equations~(\ref{eq11})--(\ref{eq12}):
\begin{eqnarray}
\dot{Z}=(\mathrm{i}\Omega-\gamma)Z+h-h^\ast(\kappa+Z^2)-\sigma^2Z(1-Z^2 -3\kappa)\,,
\label{eq17}
\\
\dot\kappa=2(\mathrm{i}\Omega-\gamma)\kappa -4h^\ast Z\kappa -\sigma^2[4\kappa(1-3Z^2)-(1-Z^2)^2]\,.
\label{eq18}
\end{eqnarray}
\end{itemize}

\begin{figure}[t]
\center{
 \includegraphics[width=0.50\textwidth]%
 {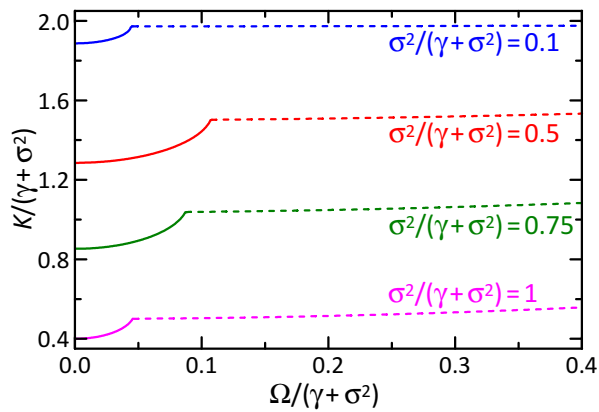}
}
\caption{Above the threshold the asynchronous state is unstable and a collective mode emerges; at low frequency the instability is monotonous (solid lines), at moderate and high frequencies the instability is oscillatory (dashed lines, equation~(\ref{eq22}))}
\label{fig1}
\end{figure}

\subsection{Linear stability of asynchronous state}
The population~(\ref{eq01}) with multiplicative intrinsic noise cannot maintain a uniform distribution of phases; equation system~(\ref{eq11})--(\ref{eq13}) does not admit the solution $\kappa_1=\kappa_{n>1}=0$. However, the system admits solutions with $\kappa_1=0$, where the Kuramoto coupling~(\ref{eq14}) between population elements also vanishes. Hence, we consider the state of $\kappa_1=0$ with no interaction between oscillators as an asynchronous one. System~(\ref{eq11})--(\ref{eq13}) also yields all odd $\kappa_{2l+1}=0$ for this asynchronous state.

Linearization of~(\ref{eq17}) determines the stability properties of the asynchronous state:
\begin{equation}
\dot{Z}=(\mathrm{i}\Omega-\gamma)Z+\frac{K}{2}(Z-Z^\ast\kappa)-\sigma^2Z(1 -3\kappa)\,,
\label{eq19}
\end{equation}
where a time-independent second cumulant $\kappa$ is given by (\ref{eq16}) and (\ref{eq18}) identically:
\begin{equation}
\kappa=\frac{\sigma^2}{4\sigma^2+2\gamma-\mathrm{i}2\Omega}\,.
\label{eq20}
\end{equation}
With (\ref{eq19}) and (\ref{eq20}), one can find the exponential growth rates of linear perturbations:
\begin{equation}
\lambda=-\gamma-\sigma^2(1 -3\mathrm{Re}\kappa)+\frac{K}{2} \pm\sqrt{\frac{K^2\sigma^4/16}{\Omega^2+(\gamma+2\sigma^2)^2}-\left(\Omega +3\sigma^2\mathrm{Im}\kappa\right)^2}\,,
\label{eq21}
\end{equation}
where $\kappa=\frac{\sigma^2}{2}\frac{\gamma+2\sigma^2+\mathrm{i}\Omega}{\Omega^2+(\gamma+2\sigma^2)^2}$\,.

\begin{figure}[t]
\center{
 \includegraphics[width=0.32\textwidth]%
 {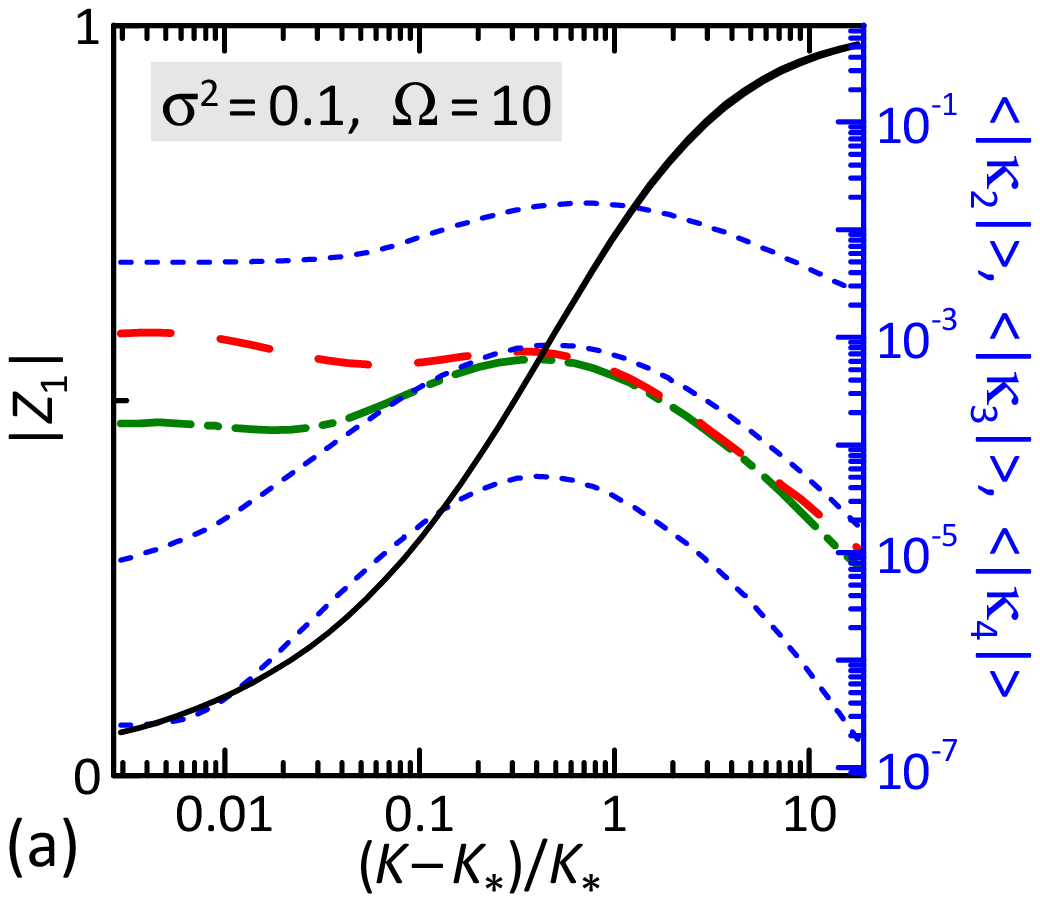}\;\;
 \includegraphics[width=0.32\textwidth]%
 {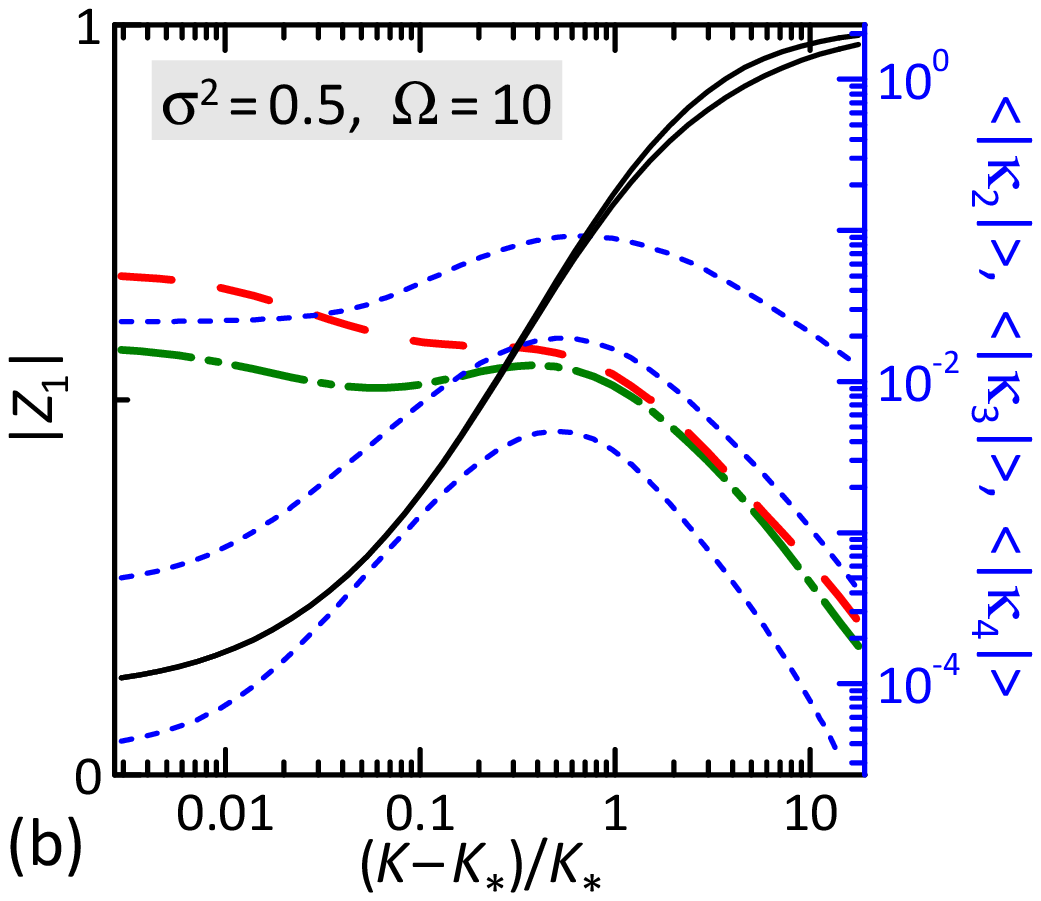}\;\;
 \includegraphics[width=0.32\textwidth]%
 {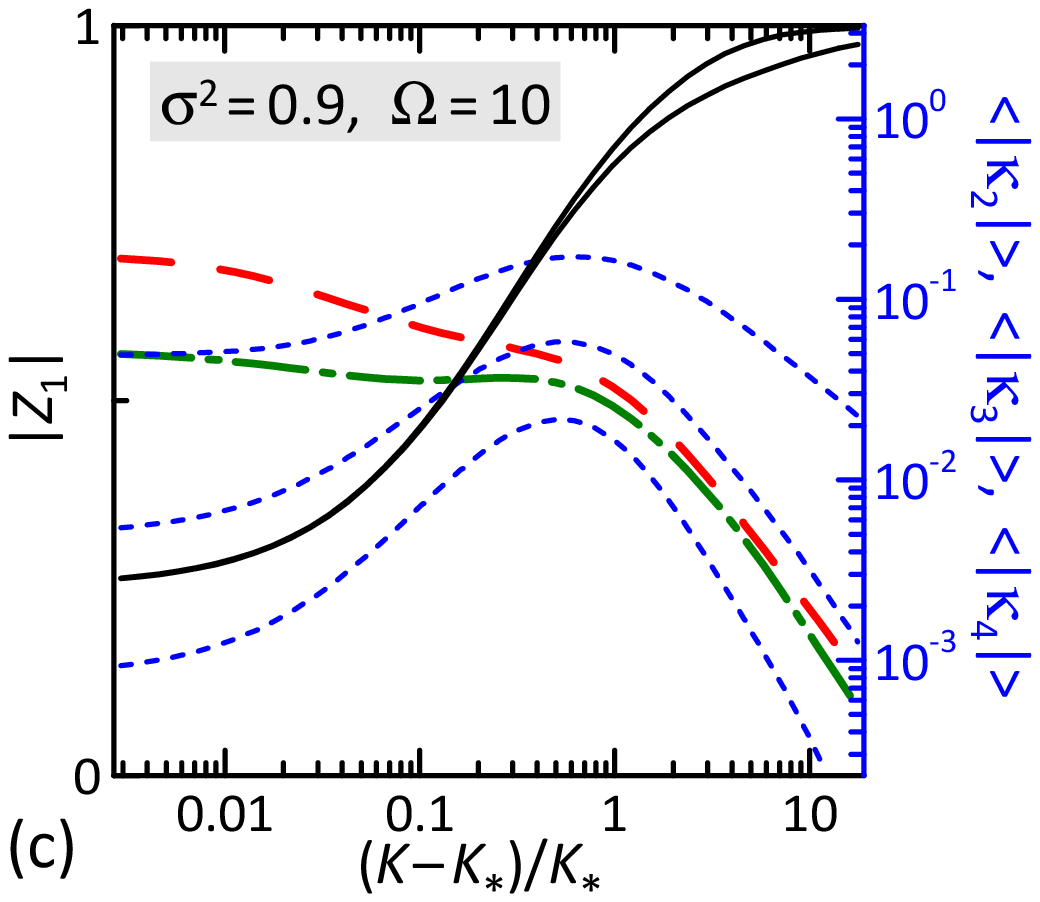}\\[10pt]
 \includegraphics[width=0.32\textwidth]%
 {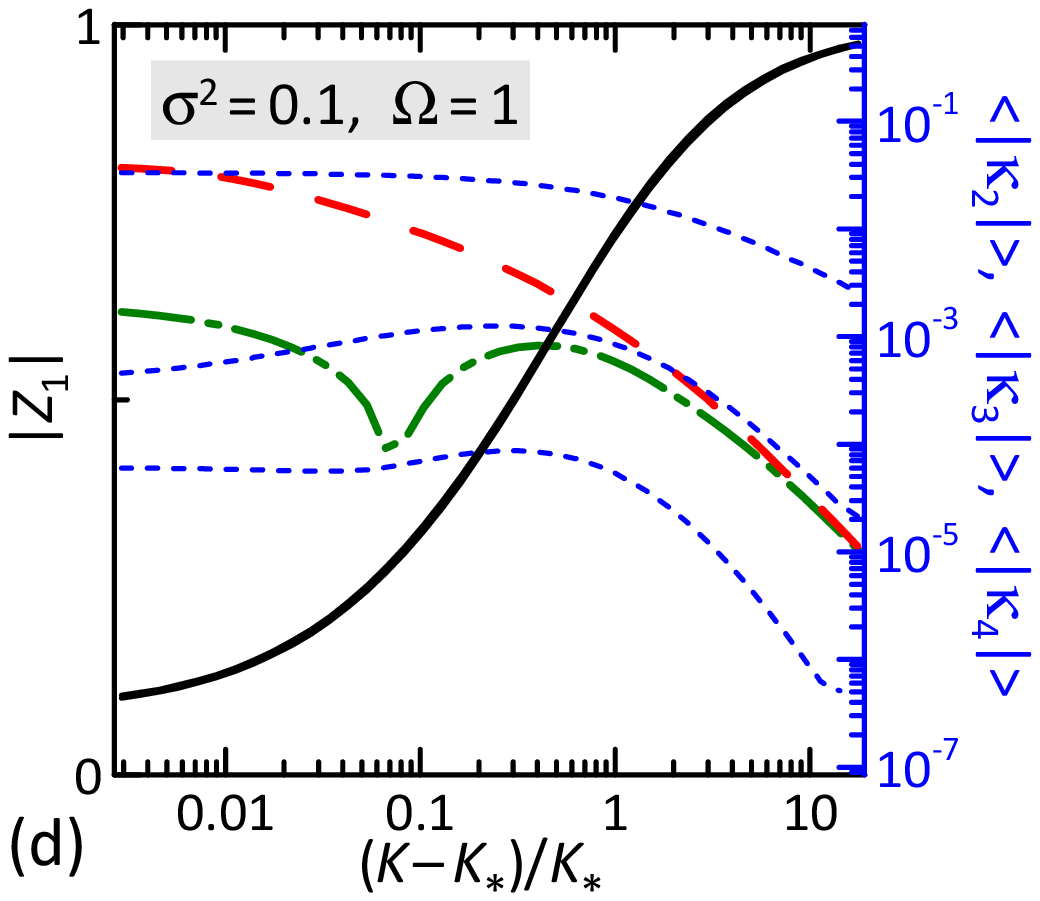}\;\;
 \includegraphics[width=0.32\textwidth]%
 {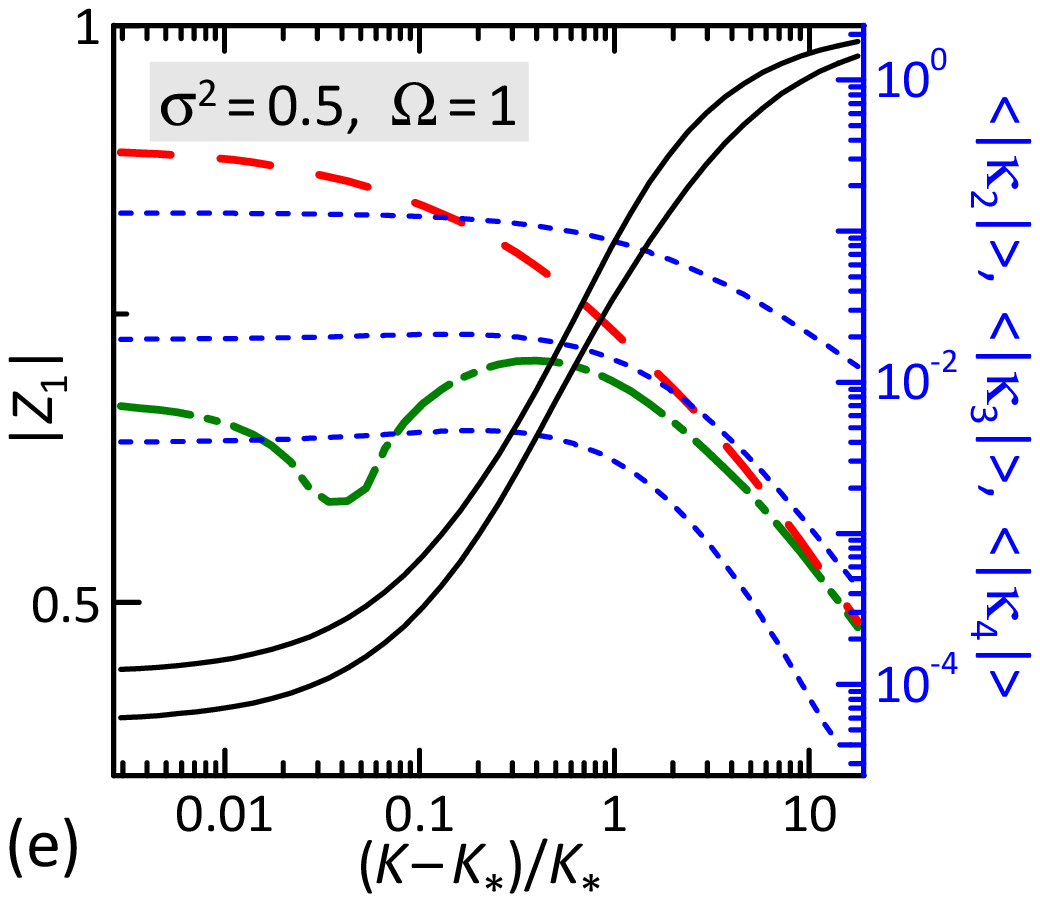}\;\;
 \includegraphics[width=0.32\textwidth]%
 {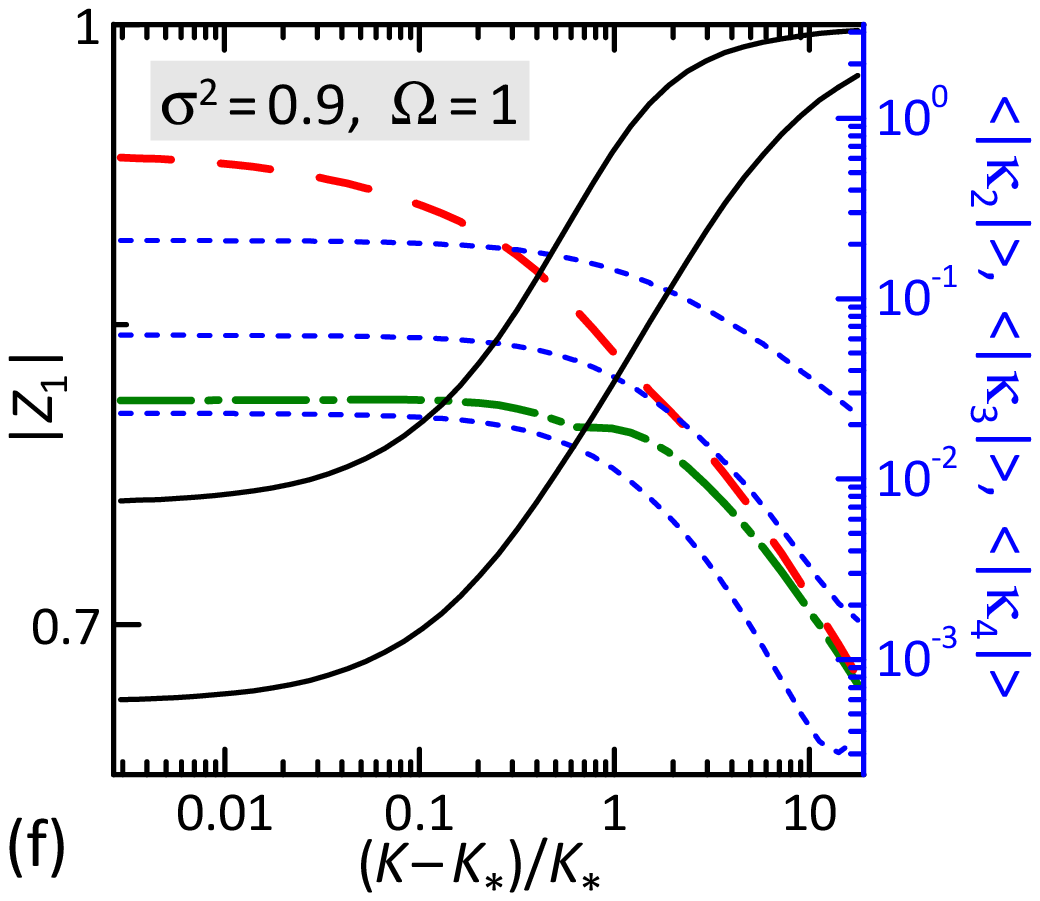}\\[10pt]
  \includegraphics[width=0.32\textwidth]%
 {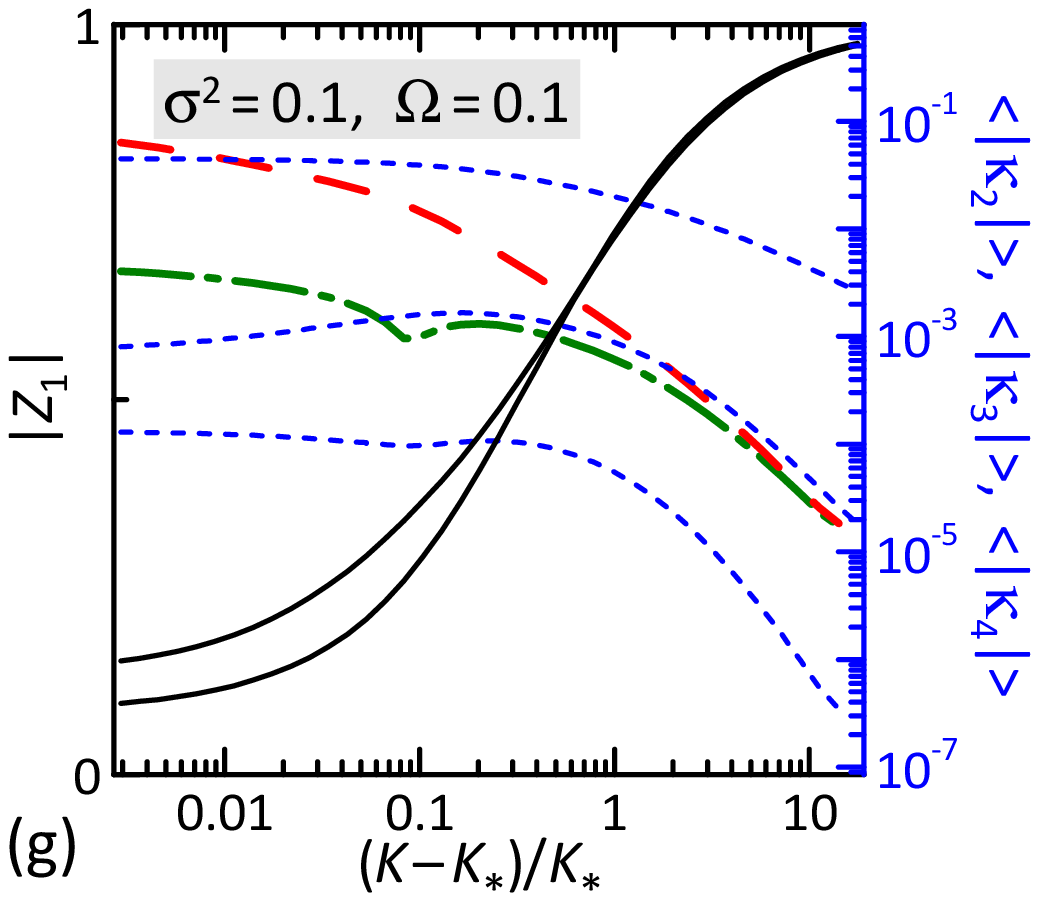}\;\;
 \includegraphics[width=0.32\textwidth]%
 {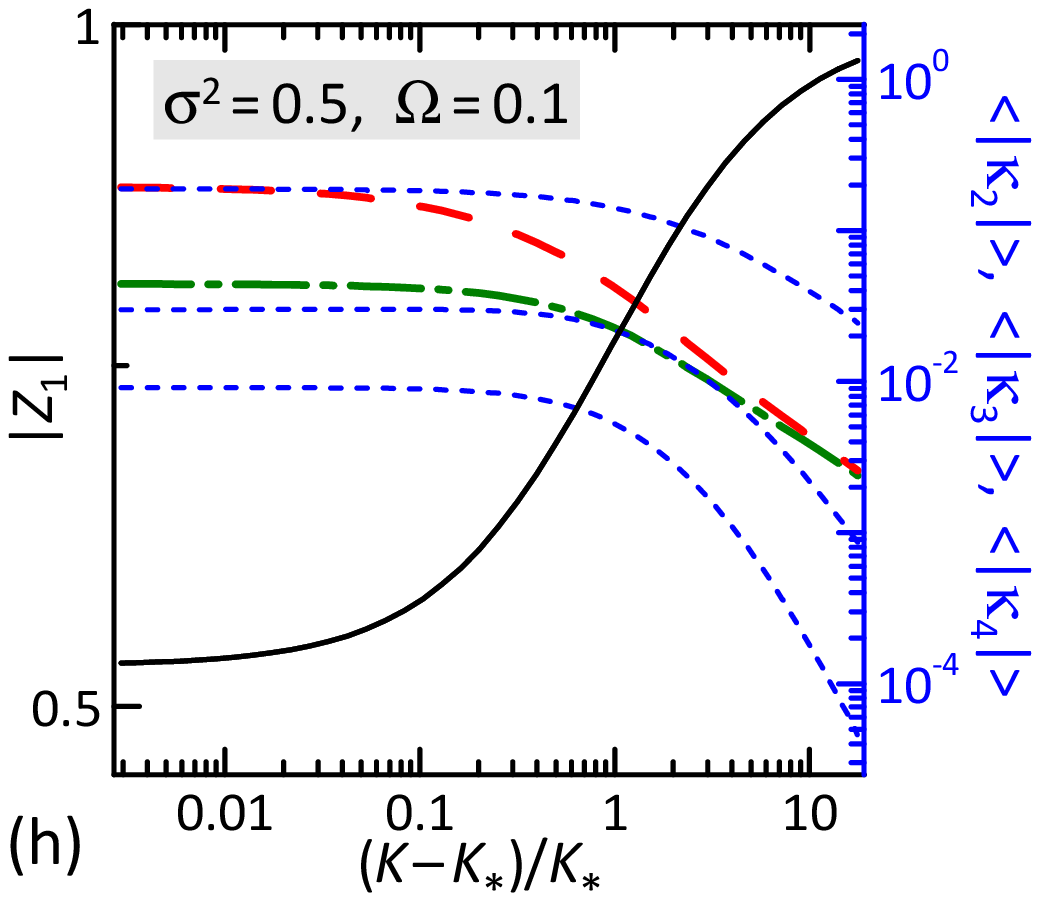}\;\;
 \includegraphics[width=0.32\textwidth]%
 {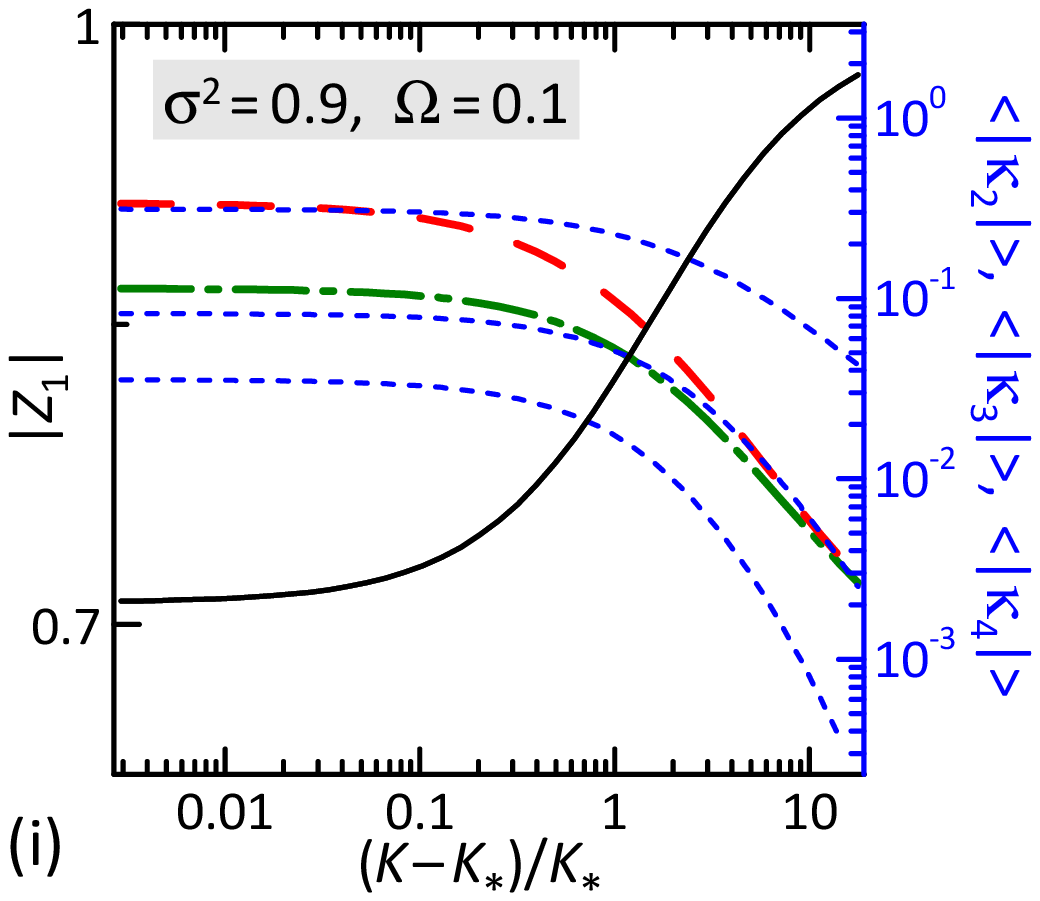}
 }
\caption{The maximal and minimal values of $|Z_1|$ are plotted for `exact' stable solutions with the black solid lines; these lines merge for time-independent states in panels (h,i), in all other panels they may be close but always do not coincide.
The circular cumulants of orders 2--4 (from top to bottom) for exact solutions are plotted with blue short-dashed lines (see the right axes for scale). Red long-dashed lines: the error of the C20 reduction (for the oscillatory regimes, the error is calculated as the half-sum of the distances between the Poincar\'e section points for hyperplanes $\mathrm{Re}Z=0$ and $\mathrm{Im}Z=0$; see the right axes for the error scale), green dash-dotted lines: error of the C2e reduction.}
\label{fig2}
\end{figure}

For sufficiently large frequency $\Omega$ the growth rate is complex-valued and the instability is oscillatory, with the threshold
\begin{equation}
K_\mathrm{osc}=2(\gamma+\sigma^2)
 -\frac{3(\gamma+2\sigma^2)\sigma^4}{\Omega^2+(\gamma+2\sigma^2)^2}\,.
\label{eq22}
\end{equation}
For small $\Omega$ the instability becomes monotonous. The expression for the monotonous instability boundary is too lengthy and we present only the results of calculations with equation~(\ref{eq21}) in figure~\ref{fig1}.

In contrast to the additive noise case, a multiplicative noise breaks the rotation symmetry of the system, and can create non-rotating macroscopic states which emerge via a monotonous instability of the asynchronous state.
For fixed~$\gamma$, multiplicative noise always uplifts the threshold for both the oscillatory and monotonous instability.

\section{Results of numerical simulation}
One can see, that in the high-frequency limit $\Omega\to\infty$, equation~(\ref{eq22}) expectedly yields the same instability threshold $2(\gamma+\sigma^2)$, as the case of additive intrinsic noise~\cite{OA-2008}. It is convenient to choose $(\gamma+\sigma^2)^{-1}$ as the time unit and accordingly rescale all parameters, which is equivalent to setting $\gamma+\sigma^2=1$ with $0\le\sigma^2\le1$ in (\ref{eq11})--(\ref{eq13}). The high-frequency instability threshold of the asynchronous state reads $K_\ast=2(\gamma+\sigma^2)=2$.

In figure~\ref{fig2}, the results of numerical simulation with C20 and C2e reductions [(\ref{eq15})--(\ref{eq16}) and (\ref{eq17})--(\ref{eq18}), respectively] are compared against the `exact' numerical solutions of equation chain~(\ref{eq05}) with 250 terms $Z_m$. With blue dashed lines one can observe circular cumulants $\kappa_2$, $\kappa_3$, $\kappa_4$ for the exact solutions, which indicate the relevance of approximations with finite number of cumulants. In the extreme case of the intrinsic-noise-dominated system with weak impact of all other factors (panels (f) and (i)), the minimalistic model reduction (C20) is inaccurate close to the excitation threshold of collective mode, $K\sim K_\ast$; otherwise, both two-cumulant reductions provide a decent accuracy of the representation of macroscopic collective behavior of the population.

\section{Conclusion}
We have constructed a circular cumulant description for large populations of Ott--Antonsen-type phase elements subject to multiplicative intrinsic noise. For the case of a multiplicative noise, the derivation of cumulant equations is significantly more sophisticated than that for the case of an additive noise~\cite{Tyulkina-etal-2018}; meanwhile in many application the intrinsic noise is essentially multiplicative.
Two two-cumulant model reductions [(\ref{eq15})--(\ref{eq16}) and (\ref{eq17})--(\ref{eq18})] are suggested on the basis of the infinite cumulant equation chain~(\ref{eq10}) and analyzed.

For a phase oscillator population with Kuramoto global coupling, the case of a multiplicative noise converges to the case of an additive one only in the high-frequency limit. Moreover, for low frequencies, the instability of the asynchronous state to formation of a macroscopic collective mode becomes monotonous. Two-cumulant model reductions provide a reasonable accuracy for the macroscopic description of the population dynamics. Meanwhile, the Ott--Antonsen ansatz and the Gaussian approximation fail to represent the system dynamics accurately for non-high frequencies.

Presented results may have important implications for networks of quadratic integrate-and-fire neurons, where an effective intrinsic noise emerges as additive for the membrane voltage~\cite{Ulner-etal-2018,Volo-Torcini-2018}, but is multiplicative for the phase variable. For a population of identical neurons, in some cases, one can expect the vanishing intrinsic noise to force a population away from the Ott--Antonsen manifold.
As a result, even for a weak intrinsic (endogenous) noise, one can expect qualitatively new regimes of macroscopic behavior of population, which are not possible near the Ott--Antonsen manifold~\cite{Volo-Torcini-2018,Ratas-Piragas-2019}.

\section*{Acknowledgements}
The authors thank L.~S.\ Klimenko for useful discussions.
The work was supported by the Russian Science Foundation (grant no.\ 19-42-04120).

\section*{References}

\bibliographystyle{elsarticle-num}

\end{document}